# Nanojet Visualization and Dark-field Imaging of Optically Trapped Vaterite Capsules with Endoscopic Illumination

Andrei Ushkov[1,2]*, Andrey Machnev[1,2], Denis Kolchanov[1,2], Toms Salgals[3,4], Janis Alnis[5], Vjaceslavs Bobrovs[3], Pavel Ginzburg[1,2]

[1]Department of Electrical Engineering, Tel Aviv University, Ramat Aviv, Tel Aviv 69978, Israel
[2]Light-Matter Interaction Centre, Tel Aviv University, Tel Aviv, 69978, Israel
[3]Institute of Telecommunications, Riga Technical University, 12 Azenes Street, 1048 Riga, Latvia
[4]Nanophotonics research laboratory (NANO-Photon Lab.), Riga Technical University, Azenes street 12, LV-1048, Riga, Latvia
[5]Institute of Atomic Physics and Spectroscopy, University of Latvia, 3 Jelgavas Street, 1004 Riga, Latvia

**Abstract**

Optical responsivity grants biomedical capsules additional capabilities, promoting them towards multifunctional theragnostic nanodevices. In this endeavor, screening candidates under conditions that closely resemble in situ environments is crucial for both the initial optimization and the subsequent inspection stages of development and operation. Optical tweezers equipped with dark-field spectroscopy are among the preferable tools for nanoparticle imaging and refractometry. However, the effectiveness of conventional illumination and light collection arrangements for inspecting anisotropic complex inner composition particles is quite limited due to reduced collection angles, which can result in the omission of features in scattering diagrams. Here we introduce an endoscopic dark-field illumination scheme, where light is launched on an optically trapped particle from a single-mode fiber, immersed into a fluid cell. This arrangement disentangles illumination and collection paths, thus allowing the collection of scattered light with a very high numerical aperture. This methodology is applied to vaterite nanocapsules, which are known to possess strong anisotropic responses. Tweezer configuration allows revealing optical properties for different crystallographic orientations of vaterite, which is complex to do otherwise. Furthermore, endoscopic dark-field images reveal the emergence of polarization-dependent long-range photonic nanojets, which are capable of interacting with nearby particles, demonstrating a new pathway for nanojet image formation.

*corresponding author

# 1. Introduction

Optically responsive capsules find use in various drug delivery scenarios, where a high level of targeting, in situ imaging capabilities, and controlled release alongside a remote activation are among the properties demanded for implementing frontier theranostic devices[1,2]. Significant preliminary efforts must be devoted to optimizing and characterizing particles at the initial design stage. Single particle characterization offers more precise and detailed information about individual particle properties[3], leading to a better understanding of their behavior and performance, compared to bulk assembly characterization which may mask individual variations and unique characteristics[4,5]. The ability to screen individual particles enables the construction of more accurate statistics and accounts for side effects, which might otherwise be overlooked in assembly studies[6]. In these endeavors, optical tweezers offer a versatile solution, enabling the trapping of a single particle while keeping the flexibility in characterization methods[7,8] and allowing for a high degree of automation [9,10]. Propulsion and sorting capabilities can be attained using adaptive optics (such as holographic tweezers [11], diffractive structures [12], or setups with steerable mirrors [13]) when integrated with a processing unit in a closed feedback loop or manually [14,15]. Optical tweezers enable the noninvasive manipulation of particles in natural biological environments, such as cell cultures, thereby reducing the constraints and limitations associated with the use of substrates [16]. After trapping, nanoparticles can be subjected to spectroscopic measurements. While fluorescent imaging remains a prevalent technique in bioimaging, label-free spectroscopy offers the advantage of extracting essential optical properties of particles, including detailed information on their refractive index tensor, without the need for labeling [17,18]. Refractometry allows deducing information on the internal particle properties and anisotropy, given the structure is birefringent. The latter property is beneficial for cross-polarization imaging and for inducing optomechanical particle rotation [19]. The rotary degree of freedom can be utilized for fluid mixing and for controlling particle dissolution rates among other possible applications [9]. All those factors facilitate a need to perform quantitative spectroscopy of optically trapped nanoparticles. The imaging methodology will be outlined first, followed by its application to a drug delivery-relevant vaterite nanoparticle, a leading candidate for creating optically-responsive drug delivery capsules.

Direct imaging of particles with bright-field techniques is diffraction-limited and thus is typically restricted to investigating larger-scale entities, spanning several microns in size [20]. On the other hand, dark-field spectroscopy allows capturing the optical properties of nanoscale particles by assessing their scattering field effectively isolating it from the illumination background [21]. Imaging of silver nanoparticles, as small as 20 nm in radius was reported via hyperspectral darkfield microscopy[22], while 5 nm particles can be also optically trapped in a gap of plasmonic nanoantenna and monitored via the nonlinear optical signal[23]. Very small (5.4 nm in diameter) gold nanoparticles were managed to optically trap for a short time (2-3 s) in a single-beam optical trap[24]. However, implementing dark-field spectroscopy with optical tweezers is far from being a trivial task. Existing implementations employ back focal plane (Fourier) filtering to differentiate between illumination and collection angles, thereby maintaining a dark background. To reveal the complexity of implementing this technique within optical tweezers, dark-field imaging strategies will be revised next. Figure 1 surveys the main techniques, shown here in the order of increasing the dark field illuminating incidence angle (from (a) to (d)).

The most commonly used dark field arrangement, shown in Fig.1b [25], is widely implemented on commercially available microscopes. Incident light is passed through a darkfield filter cube to achieve grazing angle illumination. The scattered light is collected through a complementary spatial filter. The main disadvantage of this though very practically useful implementation is the comparatively low numerical apertures (NA) (~0.9) in imaging, in contrast to the oil objectives. The problem of the limited collection NA is partially solved in a vertical illumination setup (Fig. 1a), where only close-to-vertical angles are "reserved" for illumination. Thus, the high-NA oil objectives can be used, which improves the resolution and image contrast. Being convenient for performing standard assessments, the previously mentioned techniques cannot recover the entire scattering diagram, as they block certain collection angles, reserved for the illumination. For example, the investigation of scattering patterns of plasmonic antennas [26,27,28,6] and photonic nanojets [29,30] demands collecting light from a broad range of angles. This issue is addressed by physical decoupling of the illumination and scattering collection paths (Fig.1c) [31], which unblocks highly inclined incident angles appropriate for both dark field measurements. The main complexity of this appealing scheme comes from a need to align the setup not only along optical axes. Another practical limitation arises from the form factor of long working distance objectives, which often presents challenges in integrating into the setup. All those upright microscope configurations can be upgraded with optical tweezers, given a high-NA immersion objective is introduced underneath the sample (inverted microscope scheme). However, a critical limitation of this seemingly advantageous setup is the optical thickness of the fluid cell where the trapped particle is located. In this scenario, at least one of the objectives must focus light through a fluid volume, unless the sample is thin, thus presenting an additional experimental challenge. While all the previously mentioned methods can be successfully implemented alongside their constraints, hereinafter we demonstrate a methodology that aims to bypass several limitations.

Here we develop a dark field spectroscopy scheme, integrated within the optical tweezer setup to resolve both alignment and collection cone issues (Fig.1d). The illumination scheme here is different and is achieved with a bent optical fiber, immersed into the colloidal solution. Therefore, the light directly impinges on the pre-selected particle, trapped with the tweezer. This arrangement allows direct white light illumination at ultra-high incidence angles (~90°). Moreover, the optical trap keeps the particle well above the surface (hundreds of microns distance), thus disentangling it both from optical and electrostatic interactions with the sample. Furthermore, possible substrate impact on biological species is eliminated. It enables the deterministic alignment of anisotropic nanoparticles relative to optical access and the visualization of optical nanojets, which is challenging otherwise. It is also worth mentioning that the interpretation of darkfield spectroscopy results, considering optically anisotropic particles (i.e., vaterite spherulites), becomes challenging. The difficulty with standard methodologies (Fig. 1(a, b)) arises from the excitation of additional dipole components owing to a nontrivial polarization state of the illumination cone. The endoscopic illumination in combination with optical tweezer is free of such difficulties, as it ensures a proper particle self-orientation in an optical trap and a fixed incident polarization. The developed optical tool is transformative and can be applied for the inspection of biological species (i.e., bacteria) [16], cell imaging[32], and light-sheet microscopy[33]. Considering darkfield capabilities in imaging nanoscale particles[22], the tool can be applied for the examination of a wide range of inorganic object sizes *in situ*. Finally, the setup allows for the direct measurement of rotation frequencies in trapped anisotropic particles through nanojet 'blinking', instead of the traditional quadrant photodiode (QPD) technique[34]. Worth noting that mapping of photonic nanojets was previously done via postprocessing, using z-stuck images acquired by a fast-scanning confocal microscope[35], or via the nanojet observation in a highly scattering medium[36]. Furthermore, the effect of photonic nanojets bending (photonic hook) was proposed and observed in a microwave emulation experiment via scanning-probe imaging in geometrically asymmetric structures[37]. In sharp contrast to previous studies, the direct real-time observation of nanojet bending for a spherically symmetric microparticle in optically transparent non-scattering medium is reported.

The manuscript is organized as follows: the experimental setup is described first and then used to acquire dark-field spectra from optically trapped anisotropic vaterite particles. Dark-field images revealed the emergence of polarization-dependent long-range photonic nanojets, which are capable of interacting with nearby particles. This phenomenon is comprehensively studied and verified with a full-wave numerical model, demonstrating a new pathway for nanojet image formation.

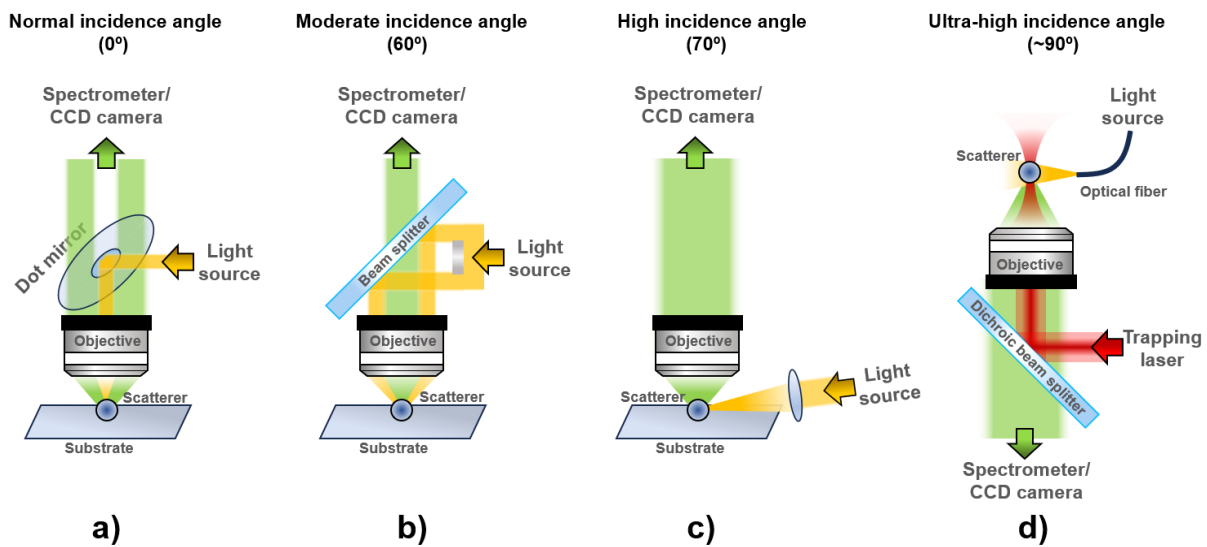

**Figure 1. (a)-(c) Examples of existing dark field principal schemes with different illuminating light incidence angles. (d) Endoscopic illumination scheme. The setups from a) to d) are sorted in ascending order of the incidence angle.**

## 2. Results

### 2.1 Vaterite particle

Vaterite nanoparticles were selected as a case study for applying our methodology to entities relevant to the field of biomedicine. Considerable efforts are dedicated to the development of particles for drug delivery. Broadly, these approaches can be categorized as organic-based or inorganic, each with numerous subsections for classifying the carrier's designs. Inorganic approaches might be preferable for implementing optically-responsive functionalities and thus will be favored here over their counterparts. The calcium carbonate ($CaCO_3$) nanoparticles have three crystalline forms- vaterite, calcite, and aragonite. Vaterite, a metastable polycrystal also termed spherulite, features high porosity, facilitating a high payload capacity, and is noteworthy for its practical, straightforward, and low-cost self-assembly synthesis. All of these indicate its potential as a biocompatible container for targeted delivery of therapeutic compounds into living cells and tissues. Vaterite particles also possess a moderately high optical anisotropy (Δn≈0.1 in the visible range), which promotes using them in microrheology and microfluidics. For example, micron-scale particles rotate under a circularly polarized illumination and can attain GHz angular velocity, being trapped in a vacuum[38]. Thereby, vaterite encompasses a great variety of promising photonics-based applications alongside its clear biomedical relevance. Considering chemical self-assembly, leading to not always deterministic outcomes, our goal is to conduct near-to-in-situ inspections of optically responsive capsules. In this paper, we use two diameters $D_{particle}$ of vaterite particles: ~1 micron and ~3 micron. The 1μm particles are employed in the dark-field spectroscopy (see Section 2.3), whereas 3μm particles are used for the nanojets observation (see Section 2.4). In addition, we demonstrate the darkfield imaging capabilities of our setup for vaterite particles down to 250 nm in radius (see Supplementary Materials).

## 2.2 Experimental Setup

The basic underpinning idea of the measurement principle appears in Fig. 2(a), demonstrating an optically trapped particle, locally illuminated with a single-mode fiber approaching the particle at ~90° angle. The scattered field is collected with the same objective (Olympus UPlanFL N 100x Oil, NA 1.30), which is used for trapping. The schematics of the optical setup appear in Fig. 2(b). The trapping diode laser (980 nm wavelength, output power 300 mW) is launched through λ/2 and λ/4 waveplates for polarization control. The vaterite particle (elaborated on below, SEM image in Fig. 2(a), inset) is trapped in a confocal dish filled with ethanol to prevent particle dissolution. After being scattered on the optically immobilized particle, the trapping laser signal is collected by 20x objective (Mitutoyo M Plan Apo NIR 20x Dry, NA 0.4) for further focusing on the quadrant photodiode (QPD). The QPD, commonly used for trapped particle rotation measurements, Brownian motion tracking, and trap stiffness calibration, is mounted at the upper arm of the confocal scheme[34]. To illuminate the microscopic field of view, a white mounted LED (Thorlabs MCWHL5) is used as a bright-field source (it can be switched off to get a better contrast for nanojet imaging). A supercontinuum laser (YSL Photonics SC-PRO) is used for dark-field illumination through a single-mode fiber. The optical fiber is mounted on a piezo XYZ stage to control the fiber tip. A micromanipulator (Scientifica PathStar), granting a micron-scale precision, is used for this purpose (Fig. 2(c)). A single-mode fiber SMF patch-cable compliant with the ITU-T G.652 standard was used for the white light illumination. In the fiber preparation process, the end of the patch cable was cleaved at 90°, and then, to get a constant bending angle, the prepared fiber was inserted into the $SiO_2$ capillary (outer diameter 0.8mm). The burning temperature of the propane-butane gas flame from the micro torch melted and softened the capillary with fiber slowly within the flame touch point. As a result, the capillary end with a fiber output was bent at an angle of ~45° with respect to the rest of the capillary tube, which was mounted on the micromanipulator.

Although the micromanipulator-controlled fiber movements create liquid flows in the vicinity of the trapped particle, they do not sweep trapped particles away due to the relatively slow speed (~1 mm/s). The light from both the dark- and bright-field sources, interacting with the particle, is collected with the high-NA objective (Olympus UPlanFL N 100x Oil, NA 1.30) and goes to the spectrophotometer (Avantes AvaSpec ULS2048L) and CCD camera (Teledyne FLIR Grasshopper3). For an accurate dark-field acquisition the bright-field source is switched off. Due to the endoscopic fiber illumination, a full NA=1.30 of the microscope objective is used for the signal collection. It should be noted, that in the case of the hollow-cone darkfield illumination by the same objective, a smaller effective $NA_{collecting}$≈1.1 would be used for darkfield imaging, as the rest should be reserved for the illumination channel. In addition, a hollow-cone illumination leads to the averaging of scattering diagrams of multiple electric dipoles, excited from different azimuthal orientations. Taking these two factors into account, we estimated a twofold increase in signal-to-noise ratio in our setup (see Supplementary Materials).

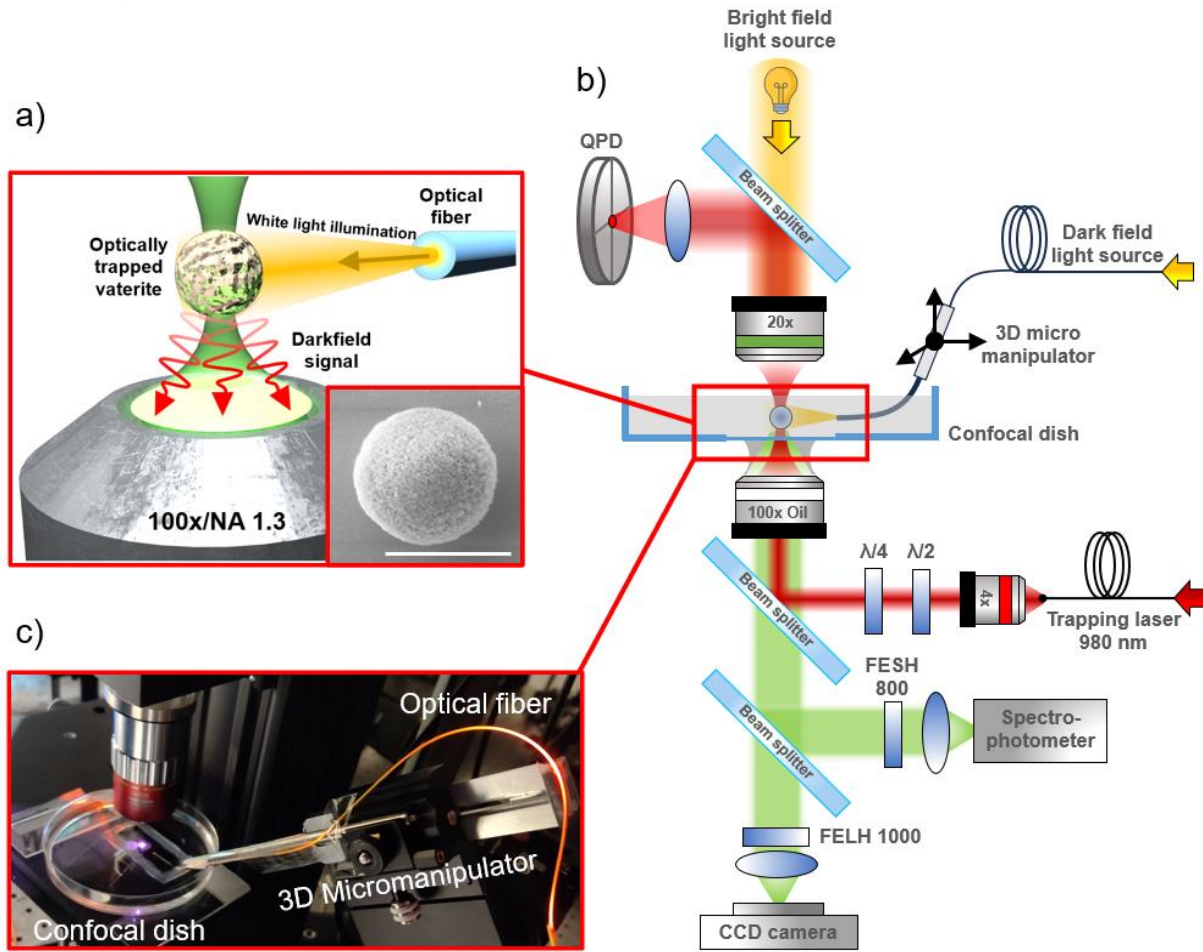

**Figure 2. Optical tweezer-based dark-field setup with endoscopic illumination. (a) The basic principle behind the dark-field spectroscopy of locally illuminated particles. The inset shows a SEM photograph of a typical ≈1 μm diameter vaterite, the white scale bar denotes 1 μm. (b) Schematics of the experimental setup. (c) Photograph of the experimental setup: optical fiber mounted on the 3D micromanipulator illuminates the optically trapped microparticle in the confocal dish.**

### 2.3 Dark-field spectroscopy of a trapped vaterite particle

A vaterite particle with a diameter $D_{particle}$=0.9 μm was trapped with the setup. The optical trap was characterized with the aid of QPD. Fig. 3(a) demonstrates the sum of all four quadrants, which enables revealing the trapping effect - no fluctuating signal is observed prior to the particle capture, and the particle stays stable within the trap after being captured (fluctuation statistics is steady). The inset shows the bright-field image of vaterite. The trap stiffness is characterized by analyzing the position fluctuations of a trapped particle, undergoing Brownian motion. The power spectral density of the fluctuations allows extracting the trap stiffness as: $\kappa = 2\pi f_c \gamma \approx 12 \ fN/nm$, where $\gamma = 3\pi\eta D_{particle}$ is the particle friction coefficient, $\eta$ the ethanol dynamic viscosity (1184*10$^{-6}$ Pa*s at 20°C), $D_{particle}$=0.9 μm the particle radius, $f_c$≈200 Hz is the corner frequency obtained from the fitting of the experimental Power Spectrum Density function with a Lorentzian $PSD(f) = Const/(f_c^2 + f^2)$ (PSD curve is shown in Fig. 3b)[34]. Considering the details of the implemented optical path, $12\ fN/nm$ is a relatively high measured stiffness, which allows a stable manipulation of particles, minimizing the escape probability.

After trapping the particle, the dark-field spectra were recorded. The vaterite major axis is aligned with the trapping light polarization, thus enabling to obtain the particle orientation control. The experimental dark-field setup shown in Fig.2b is designed to provide the s-polarized dark field illumination with an incidence angle of 80°. The incident wavevector $\mathbf{k}_x$ in the inset in Fig.3c shows the in-plane xy-projection of the total wavevector $\mathbf{k}$.

Figure 3c demonstrates significant differences in dark-field spectra, which strongly depend on the particle's orientation. The spectra are relatively noisy, which is the result of the instabilities caused by Brownian motion. The results can be smoothed by increasing the data acquisition time. Experimental data was fit with full-wave numerical simulation, which takes into account the internal vaterite crystallinity[39] and the illumination/collection arrangement. The details of the numerical analysis will follow in the next section.

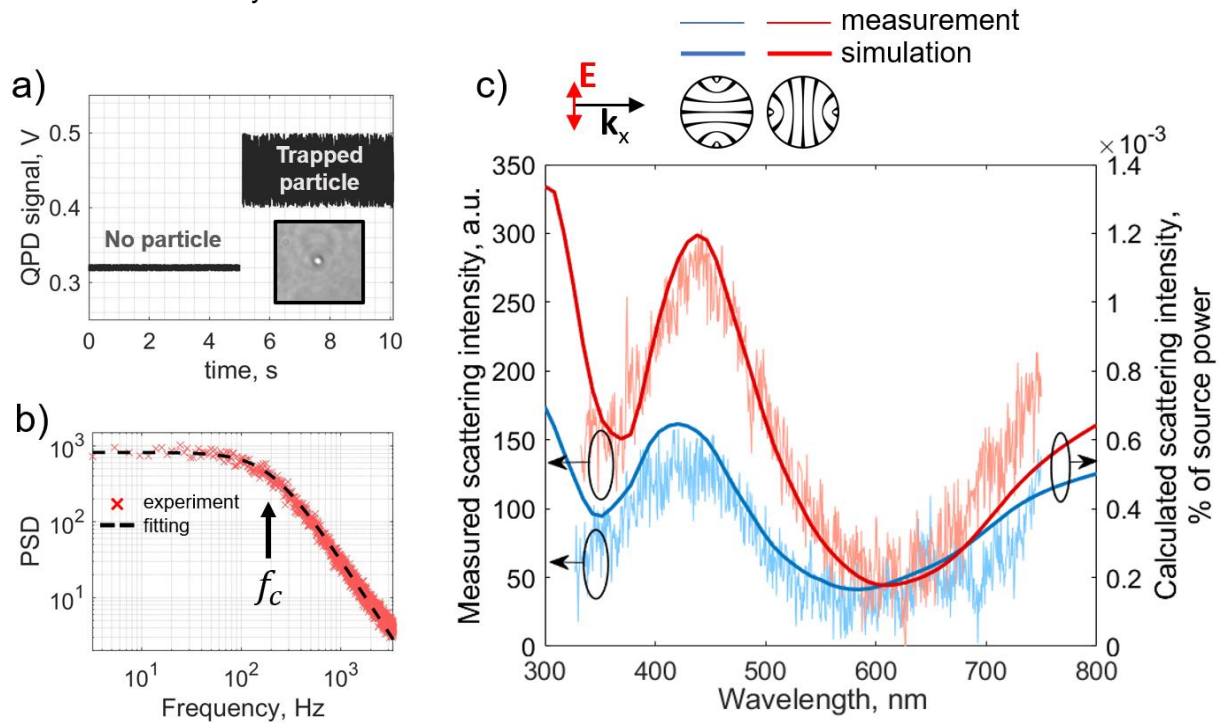

**Fig. 3. Optical trap data and dark-field spectroscopy results. (a) QPD signal (sum of 4 quadrants), demonstrating the particle trapping event. Inset - bright field microscope image of a trapped vaterite particle with 0.9 µm diameter. (b) Measured Power Spectrum Density function of the particle in the optical trap and its Lorentzian fitting with the corner frequency $f_c$=197 Hz. $f_c$ is proportional to the trap stiffness. (c) The measured and simulated dark field spectra of vaterite microparticle. Blue and red spectra correspond to different orientations of the optically anisotropic particle with respect to the incident plane wave.**

## 2.4 Nanojet blinking and inter-particle nanojets observation

During the dark-field data acquisition for vaterite size $D_{particle}$=3 micron, optical imaging was performed and revealed a set of peculiar phenomena. Specifically, a formation of nanojets was observed, which, in principle, are not expected to appear in the far-field image of the sample area.

Photonic nanojets are intense near-fields that emerge due to the interaction between light and micron-scale particles. Electromagnetic hot spots can be controlled by a particle's refractive index and shape thus tailoring near-fields quite efficiently. The nanojet bridges the gap between small-scale plasmonic nanoparticles, which offer near-field enhancement at the expense of optical loss[40], and large-scale dielectric lenses, where a wider aperture is traded for improved focusing[34]. Nanojets can enhance performance in many applications, including optomechanical manipulation, Raman spectroscopy, and bioimaging, to name just a few. In the context of developing optically responsive capsules, vaterite nanocarriers that support tunable nanojets can provide an additional advantage over low-contrast capsules, e.g., polymeric.

Nanojects, being a near-field phenomenon, require extra care in visualization. Typically, the imaging is done with either near-field (SNOM)[41], confocal scanning methods[35], or via scattering on imperfections[42]. In microfluidic applications, the jets scatter from substrates and are subsequently visualized with a microscope. In any case, an auxiliary structure is needed to map near-field properties to the far-field detection.

Here, we demonstrate that using our illumination scheme, granting ultra-high incident angles of incidence, alongside high NA oil objective, enables performing the nanojets visualization from the far field. Nanojet observation will be presented first and then followed by detailed analyses of the interaction scenarios. Figure 4 demonstrates experimental photographs in near-infra-red (NIR) of vaterite and silica microparticles with similar diameters $D_{particle}$~3 micron, generating photonic nanojets.

At first, orientation-sensitive nanojets, generated by vaterite particles, were observed (Fig. 4(a) and (b)). Vaterite, being a strongly anisotropic particle, aligns in an optical trap along the polarization of the laser. Switching the linear polarization of the trapping laser with a half-wave plate causes the rotation of the particle. This allows us to observe nanojets for two mutually orthogonal arrangements of the particle with respect to the incident (from the fiber) light. As can be seen, the nanojet intensity strongly depends on the particle's orientation, as it is also verified with the full-wave numerical model, which accounts for the vaterite internal anisotropy (Fig. 4(c) and (d)). The maximal field enhancement, as the function of the angle between the incident light polarization and the major axis of the particle, appears in Fig.4(e). A 2-fold difference is observed, nevertheless, the anisotropy $n_{eo}/n_o=1.65/1.55≈1.06$ is relatively low. If vaterite rotates in a circularly polarized optical trap, the nanojet blinks.

In the set of experiments, we also found that nanojets can mitigate intermediate-range photonic interactions between the particles. It was found that nanojets bend towards the secondary particle, as it is shown in Fig 4(f-k), at the specific spatial scale: those interactions are neither near- nor far-field, but rather intermediate on the wavelength scale. Reference measurements without the second particle present possess no bending. While the bending phenomenon is rather universal and observed for both vaterite and silica particles under similar conditions, anisotropic vaterite demonstrates more confined and stronger jets. Those differences will be revealed next.

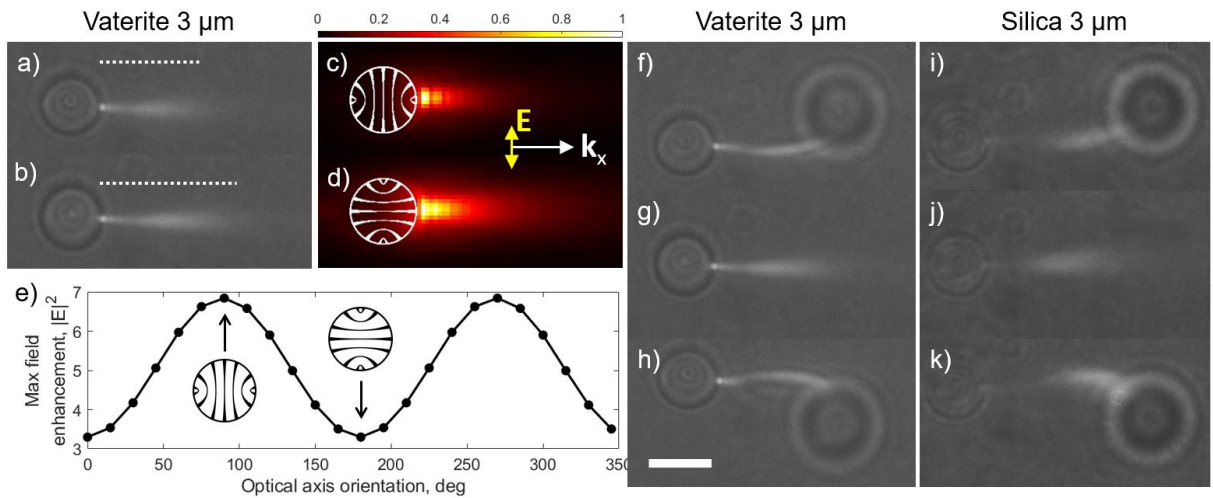

Fig. 4. Photonic nanojets from vaterite and silica microparticles with diameters of $D_{particle}$=3 µm. (a), (b) Vaterite orientation-dependent nanojets. The nanojet lengths, imaged with the CCD camera, depend on the mutual orientation between the incident field polarization and the vaterite major axis, as verified numerically - (c) and (d), respectively. Dashed white lines in (a) and (b) denote the visual impression of the corresponding nanojets lengths. The incident wavevector $k_x$ in the inset in (c)-(d) shows the horizontal xy-projection of the total wavevector k. (e) The vaterite photonic nanojet field enhancement as the function of the vaterite major axis angle with respect to the incident wave polarization. Photonic nanojets are produced by the (f-h) vaterite and (i-k) silica particles. Nanoject bending towards a nearby particle (f), (h), (i), and (k). (g) and (j) - reference jets without a secondary particle present. The photographs (a)-(b), (f)-(k) were recorded in NIR. Numerical calculations are CW at 1.1 µm. Scale bar in (h) denotes 3 µm.

## 2.5 Theory of Nanojet Image Formation

After observing the jets, the physical mechanism behind their direct visualization will be revealed. In this endeavor, we performed full wave finite difference time domain (FDTD) simulations of the particles, illuminating them with a plane wave at a high 80° incident angle. The schematic layout of the simulation appears in Fig. 5(a).

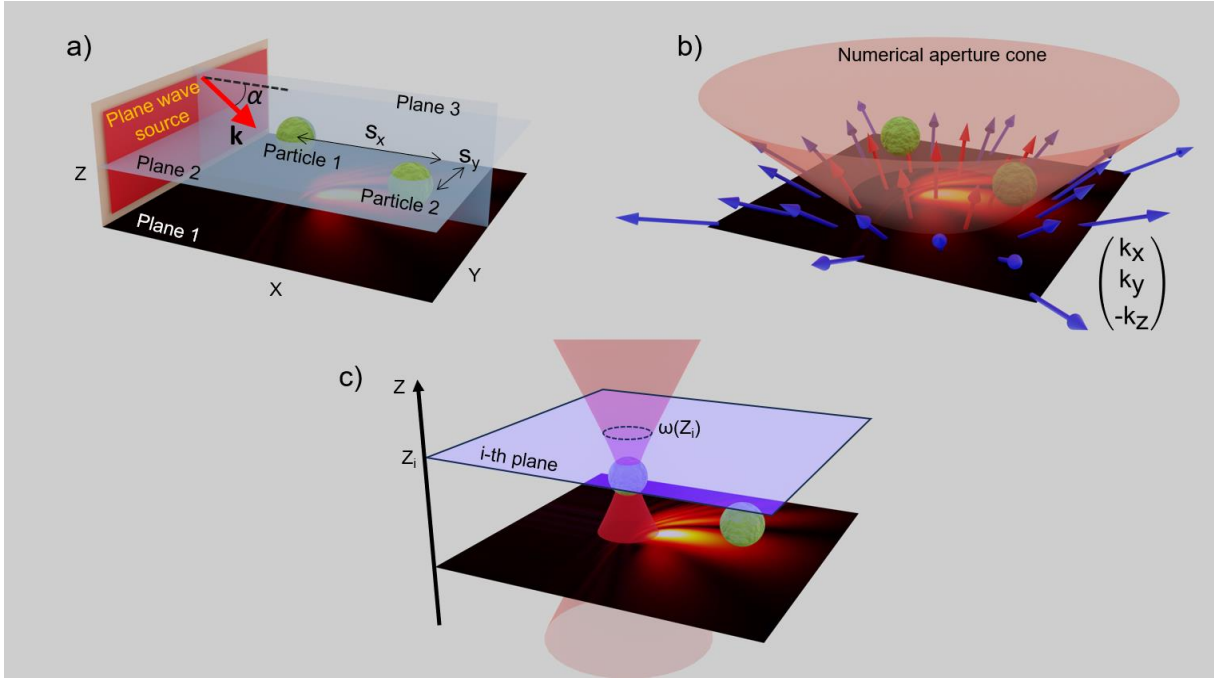

**Fig. 5. The numerical analysis – main steps. (a) The principal scheme - FDTD model with two particles. The plane wave (CW, 1.1 μm wavelength) source has an incidence angle of 90°- α = 80°. (b) The calculated electromagnetic fields in Plane 1 (located beneath the particles in isotropic medium) are used for angular spectrum decomposition and subsequent field calculations for a set of planes around the particles (i-th planes in (c)). Planes 2 and 3 in (a) are central cuts of the FDTD simulation box. To simulate the images on the CCD camera (i.e., Figs. 6(a-f)), a set of intensity distributions in i-th planes is averaged (see details in text).**

Two spherical particles with diameters $D_{particle}$=3 μm are placed at a distance $s_x$=4D and $s_y$=-2*D..2*D (varying value) from each other. Vaterite particles are modeled with their complex position-dependent anisotropic tensor, accounting for the internal crystallinity[39]. $n_{eo}$=1.65 and $n_o$=1.55 are the local principal axis refractive indexes. The principal central high-index optical axis of the particle is considered to be directed along $s_x$. For isotropic silica particles, the tabulated values were used[43]. The embedding medium is ethanol (n≈1.36 in visible[44]). The plane wave with grazing angle α=10° (and incident angle 80°) is used as a source (see Fig.5 (a)). Owing to computational capabilities, the full wave simulation allows considering relatively small computation volumes, which cannot exceed several wavelengths in each dimension. This is especially relevant to vaterite, which has to be accurately meshed to account for its internal anisotropy. However, the plane-wave decomposition method[45] allows projecting near-field distribution to the far-field propagating waves. Those can be directly linked to form a microscope image, considering the available NA.

The scattering field data $\mathbf{E}(x_n, y_m, z)$ from 'Plane 1' is used for the angular spectrum decomposition. The plane is located beneath the particles and does not intersect with dielectric structures (i.e., the refractive index at each point on the plane is uniform and isotropic. Thus, following the 2D Fourier transform on a discrete mesh[45]:

$$\hat{\mathbf{E}}(k_x, k_y, z) = \sum_{n,m} \mathbf{E}(x_n, y_m, z) \exp(-ik_x x_n - ik_y y_m) \frac{\Delta x_n \, \Delta y_m}{(2\pi)^2}, \quad (1)$$

where $(x_n, y_m)$ is the spatial coordinate mesh in 'Plane 1' and $(k_x, k_y)$ is the mesh in reciprocal space. $k_z$ is calculated from the wave dispersion within the homogeneous material (ethanol in this case).

A set of wavevectors $(k_x, k_y, k_z)$ after the decomposition is depicted by coloured arrows in Fig. 5(b). Note, that for the sake of visualization $(k_x, k_y, -k_z)$ is presented, i.e., the arrows are depicted upwards nevertheless the propagation is downwards to the objective. On this plot, red arrows denote the waves inside the microscope objective NA cone (NA=1.3, medium is ethanol), while blue arrows - outside it. Thus, only wavevectors depicted in red pass through the microscope objective and form an image on the camera.

To model the camera image, the light intensity distributions $\mathbf{E}(x_n, y_m, z_i)$ are reconstructed on a set of horizontal planes $z_i$ (called i-th planes in Fig.5c) using the free-space propagator (see Fig. 5(c)):

$$\mathbf{E}(x_n, y_m, z_i) = \sum_{k_x, k_y} \hat{\mathbf{E}}(k_x, k_y, z_i) \exp(ik_x x_n + ik_y y_m) \Delta k_x \, \Delta k_y, \qquad (2)$$

where

$$\hat{\mathbf{E}}(k_x, k_y, z_i) = \hat{\mathbf{E}}(k_x, k_y, z_{\text{Plane 1}}) \exp([ik_z(z_i - z_{\text{Plane 1}})]),$$
$$k_z = -\sqrt{\left(\frac{2\pi}{\lambda} n_{medium}\right)^2 - k_x^2 - k_y^2}, \qquad (3)$$

The "-" sign in equation (3) for $k_z$ shows the direction of scattered energy flux, opposite to the Z-axis direction.

A set of horizontal $z_i$ planes is chosen in the range $z_{\text{particle 1}}-D_{\text{particle}}/2 < z_i < z_{\text{particle 1}} + D_{\text{particle}}/2$ to cover Particle 1 (see Fig.5a), and the corresponding scattered field intensity maps $I(x_n, y_m, z_i) = |\mathbf{E}(x_n, y_m, z_i)|^2$ are calculated.

To simulate the depth of field effect, a Gaussian blur filter is applied to every map, with the standard deviation σ=ω(z)/2, where ω(z) is the Gaussian beam waist[46], see Fig.5c. The Gaussian beam Rayleigh range $z_R$ is defined by its waist radius $\omega_0$ via the objective NA: $z_R = n_{medium} \omega_0 / NA$. Finally, the simple average field intensity map is calculated over the full set of $z_i$ planes.

Figure 6 summarizes the numerical results of photonic nanojets, generated under conditions of experimental dark field setup. Colormaps show the camera images, modeled with the previously described approach. The main objective here is to compare the field intensity distribution in the plane, containing the Particle 1 and the real image, observed with the microscope objective. Figs. 6 (a and c) compare the scenarios when particles are aligned with the propagation direction of the illumination. In this case, the near-field (panel c) has fast-oscillating components, which do not make it to the camera image (panel a). It is rather expected as the fast oscillations are mapped to evanescent fields, which do not propagate to the objective. In the "lens view", the second particle appears brighter, as it scatters the jet.

Figures 6 (b and d) represent the scenario when the second particle is displaced. Here the difference between the total field in the particles' plane and the real image differ dramatically. While no bent jet appears in the near-field map, the far-field image clearly demonstrates the field attraction, which was indeed observed in the experiment.

Similar effects are predicted for silica particles, demonstrating lesser confined jets, as expected owing to a lower refractive index contrast. This modeling also confirms the experimental observation (Fig. 4).

To further investigate the bending effect and attribute it to the particle's refractive index, the following numerical study has been done. Two dielectric particles are located in correspondence with Fig.5a setup at a distance $s_x=4D_{particle}=12$ micron along X-axis and at a varying offset $s_y$ along Y-axis (see also an inset in Fig.6g). The characteristic jet bending (Max $s_y$) was calculated as a function of the particle's refractive index, see Fig.6g. The value of Max $s_y$ corresponds to the transverse offset $s_y$ when a nanojet tail intensity decays by a factor of 1/e in comparison to the aligned situation $s_y=0$ (see Supplementary Materials). It can be clearly seen that Max $s_y$ monotonically grows with the optical density. Several representative dielectric particles, made from silica, BK7, vaterite, and MR174 (a high refractive index n=1.74 optical lens material) are indicated in the inset. Overall, the results verify that the photonic nano-jet bending is the result of the near-field particles interaction, observed via the far-field mapping on the camera. The mechanism of bending observation inherently relies on the near-to-far field transformation performed by the microscope objective.

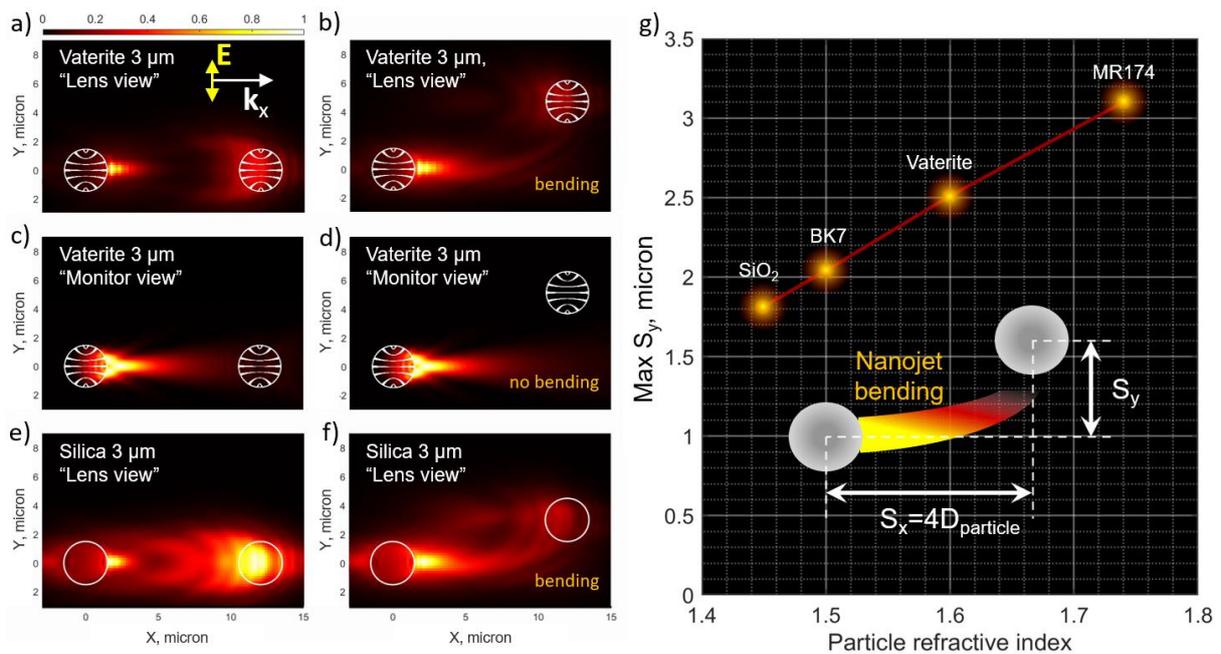

**Figure 6. Numerical analysis of inter-particle photonic nano-jet bending effect. Numerical analysis of inter-particle photonic nano-jet bending effect. (a), (b) Images of nanojets through the objective - "lens view", as elaborated in Fig.5. Vaterite particles with their central optical axes aligned with incident wavevector $k_x$ are used, as shown in the inset to (a). (c), (d) FDTD-calculated scattered near field intensity colormaps. No nanojet bending is found, in contrast to (a) and (b). (e),(f) The same as (a),(b), but with silica particles instead of vaterite ones. The particle's diameter in (a)-(f) is 3 μm. (g) Characteristic nanojet bending (Max $s_y$) as the function of the particle's refractive index. The mutual arrangement of representative particles is in the inset.**

## 3. Outlook and Conclusion

The illumination endoscopic dark-field spectroscopy method has been developed and demonstrated on several types of optically trapped particles. Optical tweezers enable the investigation, screening, and sorting of single particles in solutions based on their spectroscopic signatures. A special notion can be devoted to anisotropic and/or nonspherical particles, which can be rotated in the trap and thus investigated from several nontrivial directions. We have demonstrated this capability on vaterite nanoparticles, verifying its strong anisotropic response. Furthermore, the endoscopic architecture enables untying the spectra from substrate contributions, which might be important in several microfluidic arrangements. A distinctive advantage of this spectroscopic technique is the capability to collect scattered light from an optically trapped particle at an extremely broad range of angles. In typical widely-used arrangements, angular limitations come from a demand to grant high NA and light intensity for stable particle trapping. Furthermore, interpreting results from darkfield spectroscopy, especially when dealing with optically anisotropic particles like vaterite spherulites, presents its own set of challenges. These complexities are primarily due to the excitation of additional dipole components, which occur because of the nontrivial polarization state of the illumination cone, as illustrated in standard methodologies. The reported configuration is free of those limitations.

The endoscopic illumination allows to excite particles at enormously large grazing angles, approaching 90°. On top of the previously mentioned advantages, this arrangement enables the observation of photonic nanojets. Taking advantage of high-NA collection, we have demonstrated the direct observation of nanojets and inter-particle nanojet bending. In the latter case, the microscope image demonstrates the nanojet bending towards a nearby particle. The nature of this peculiar effect was revealed and is solely attributed to the image formation dynamics, as the near-field maps lack bending properties.

The focus of the studies was made on vaterite nanoparticles, which are among the most promising platforms for realizing theragnostic nanodevices. In this endeavor, developing optically-responsive drug capsules can grant additional advantages. The demonstrated methodology enables fast screening of capsules, thus providing a level of determinism to colloidally fabricated nanodevices.

**References**

1. Choukrani, G., Maharjan, B., Park, C. H., Kim, C. S. & Kurup Sasikala, A. R. Biocompatible superparamagnetic sub-micron vaterite particles for thermo-chemotherapy: From controlled design to in vitro anticancer synergism. Mater. Sci. Eng. C Mater. Biol. Appl. 106, 110226 (2020).

2. Calcium carbonate vaterite particles for drug delivery: Advances and challenges. Materials Today Advances 14, 100214 (2022).

3. Arita, Y. et al. Cooling the optical-spin driven limit cycle oscillations of a levitated gyroscope. Communications Physics 6, 1–7 (2023).

4. Permyakov, D. et al. Probing magnetic and electric optical responses of silicon nanoparticles. Appl. Phys. Lett. 106, 171110 (2015).

5. Koshelev, K. & Kivshar, Y. Dielectric Resonant Metaphotonics. ACS Photonics (2020) doi:10.1021/acsphotonics.0c01315.

6. Nguyen, T. M. et al. Ultralow-Loss Substrate for Nanophotonic Dark-Field Microscopy. Nano Lett. 23, 1546–1554 (2023).

7. Gilad, H. et al. Gilded vaterite optothermal transport in a bubble. Sci. Rep. 13, 1–11 (2023).

8. Samek, O., Zemánek, P., Jonáš, A. & Telle, H. H. Characterization of oil-producing microalgae using Raman spectroscopy. Laser Phys. Lett. 8, 701 (2011).

9. Ushkov, A., Machnev, A. & Ginzburg, P. Optically Controlled Dissolution Kinetics of Vaterite Microcapsules: Toward Novel Crystal Growth Strategies. Cryst. Growth Des. 23, 8009–8017 (2023).

10. Kim, K. & Park, Y. Tomographic active optical trapping of arbitrarily shaped objects by exploiting 3D refractive index maps. Nat. Commun. 8, 15340 (2017).

11. Markovich, H., Shishkin, I. I., Hendler, N. & Ginzburg, P. Optical Manipulation along an Optical Axis with a Polarization Sensitive Meta-Lens. Nano Lett. 18, 5024–5029 (2018).

12. Bogdanov, A. A., Shalin, A. S. & Ginzburg, P. Optical forces in nanorod metamaterial. Sci. Rep. 5, 15846 (2015).

13. Fällman, E. & Axner, O. Design for fully steerable dual-trap optical tweezers. Appl. Opt., AO 36, 2107–2113 (1997).

14. Kumar, A. & Bechhoefer, J. Nanoscale virtual potentials using optical tweezers. Appl. Phys. Lett. 113, 183702 (2018).

15. Landenberger, B., Yatish & Rohrbach, A. Towards non-blind optical tweezing by finding 3D refractive index changes through off-focus interferometric tracking. Nat. Commun. 12, 1–11 (2021).

16. Altindal, T., Chattopadhyay, S. & Wu, X.-L. Bacterial Chemotaxis in an Optical Trap. PLoS One 6, e18231 (2011).

17. Fiore, A., Bevilacqua, C. & Scarcelli, G. Direct Three-Dimensional Measurement of Refractive Index via Dual Photon-Phonon Scattering. Phys. Rev. Lett. 122, 103901 (2019).

18. Knöner, G., Parkin, S., Nieminen, T. A., Heckenberg, N. R. & Rubinsztein-Dunlop, H. Measurement of the Index of Refraction of Single Microparticles. Phys. Rev. Lett. 97, 157402 (2006).

19. Parkin, S. J. et al. Highly birefringent vaterite microspheres: production, characterization and applications for optical micromanipulation. Opt. Express 17, 21944–21955 (2009).

20. Wayne, R. O. Light and Video Microscopy. (Academic Press, 2013).

21. Fang, W. et al. Quantizing single-molecule surface-enhanced Raman scattering with DNA origami metamolecules. Sci Adv 5, eaau4506 (2019).

22. Théoret, T. & Wilkinson, K. J. Evaluation of enhanced darkfield microscopy and hyperspectral analysis to analyse the fate of silver nanoparticles in wastewaters. Anal. Methods 9, 3920–3928 (2017).

23. Yoon, S. J. et al. Non-fluorescent nanoscopic monitoring of a single trapped nanoparticle via nonlinear point sources. Nat. Commun. 9, 1–8 (2018).

24. Brzobohatý, O. et al. Non-spherical gold nanoparticles trapped in optical tweezers: shape matters. Opt. Express 23, 8179–8189 (2015).

25. Noskov, R. E. et al. Golden Vaterite as a Mesoscopic Metamaterial for Biophotonic Applications. Adv. Mater. 33, e2008484 (2021).

26. Jiang, W., Hu, H., Deng, Q., Zhang, S. & Xu, H. Temperature-dependent dark-field scattering of single plasmonic nanocavity. Nanophotonics 9, 3347–3356 (2020).

27. Li, G.-C., Zhang, Y.-L. & Lei, D. Y. Hybrid plasmonic gap modes in metal film-coupled dimers and their physical origins revealed by polarization resolved dark field spectroscopy. Nanoscale 8, 7119–7126 (2016).

28.     Fan, J. A. et al. Near-normal incidence dark-field microscopy: applications to nanoplasmonic spectroscopy. Nano Lett. 12, 2817–2821 (2012).

29.     Minin, I. V., Liu, C.-Y., Yang, Y.-C., Staliunas, K. & Minin, O. V. Experimental observation of flat focusing mirror based on photonic jet effect. Sci. Rep. 10, 8459 (2020).

30.     Liu, C.-Y. & Yeh, M.-J. Experimental verification of twin photonic nanojets from a dielectric microcylinder. Opt. Lett. 44, 3262–3265 (2019).

31.     Barhom, H. et al. Biological Kerker Effect Boosts Light Collection Efficiency in Plants. Nano Lett. 19, 7062–7071 (2019).

32.     Zhao, J. et al. Deep-learning cell imaging through Anderson localizing optical fiber. AP 1, 066001 (2019).

33.     Plöschner, M. et al. Multimode fibre: Light-sheet microscopy at the tip of a needle. Sci. Rep. 5, 1–7 (2015).

34.     Jones, P. H., Marag-, O. M. & Volpe, G. Optical Tweezers: Principles and Applications. (Cambridge University Press, 2015).

35.     Ferrand, P. et al. Direct imaging of photonic nanojets. Opt. Express 16, 6930–6940 (2008).

36.     Chen, X. et al. Subwavelength imaging and detection using adjustable and movable droplet microlenses. Photon. Res., PRJ 8, 225–234 (2020).

37.     Minin, I. V. et al. Experimental observation of a photonic hook. Appl. Phys. Lett. 114, 031105 (2019).

38.     Ahn, J. et al. Ultrasensitive torque detection with an optically levitated nanorotor. Nat. Nanotechnol. 15, 89–93 (2020).

39.     Noskov, R. E., Shishkin, I. I., Barhom, H. & Ginzburg, P. Non-Mie optical resonances in anisotropic biomineral nanoparticles. Nanoscale 10, 21031–21040 (2018).

40.     Li, K. et al. Balancing Near-Field Enhancement, Absorption, and Scattering for Effective Antenna–Reactor Plasmonic Photocatalysis. (2017) doi:10.1021/acs.nanolett.7b00992.

41.     Minin, I. V. et al. Plasmonic nanojet: an experimental demonstration. Opt. Lett., OL 45, 3244–3247 (2020).

42.     Characterization of photonic nanojets in dielectric microdisks. Physica E 73, 226–234 (2015).

43. Malitson, I. H. Interspecimen Comparison of the Refractive Index of Fused Silica*,†. J. Opt. Soc. Am., JOSA 55, 1205–1209 (1965).

44. Moreels, E., de Greef, C. & Finsy, R. Laser light refractometer. Appl. Opt., AO 23, 3010–3013 (1984).

45. Novotny, L. & Hecht, B. Principles of Nano-Optics. (Cambridge University Press, 2012).

46. Yariv, A. & Yeh, P. Optical Waves in Crystals: Propagation and Control of Laser Radiation. (Wiley-Interscience, 1984).

**Acknowledgments**
TAU Team acknowledges ERC StG "In Motion" (802279) and the Israel Science Foundation (ISF grant number 1115/23). RTU Team acknowledges the Latvian Council of Science project: Novel complex approach to the optical manipulation of Nanoparticles (PHOTON), No. lzp-2022/1-057. A.U. acknowledges the support of the Azrieli Foundation's Postdoctoral Fellowship.

**Contributions**
A.U. performed the experiments and analyzed the data
A.M. built the setup and helped with the experiments
D.K. provided the microparticles and performed SEM analysis
T.S. built the fiber optics part of the setup
J.A. conceived the idea, supervised the study
V.B. participated in valuable discussions
P.G. conceived the project, supervised the study, and wrote the manuscript

**Figure legends**

**Figure 1. (a)-(c) Examples of existing dark field principal schemes with different illuminating light incidence angles. (d) Principal scheme considered in this study.**

**Figure 2. Optical tweezer-based dark-field setup. (a) Basic principle behind the dark-field spectroscopy of locally illuminated particles. The inset shows a SEM photograph of a typical ≈1 μm diameter vaterite, the white scale bar denotes 1 μm. (b) Schematics of the experimental setup. (c) Photograph of the experimental setup: optical fiber mounted on the 3D micromanipulator illuminates the optically trapped microparticle in the confocal dish.**

**Fig. 3. Optical trap data and dark-field spectroscopy results. (a) QPD signal (sum of 4 quadrants), demonstrating the particle trapping event. Inset - bright field microscope image of a trapped vaterite particle with 0.9 μm diameter. (b) Measured Power Spectrum Density function of the particle in the optical trap and its Lorentzian fitting with the corner frequency $f_c$=197 Hz. $f_c$ is proportional to the trap stiffness. (c) The measured and simulated dark field spectra of vaterite microparticle. Blue and red spectra**

correspond to different orientations of the optically anisotropic particle with respect to the incident plane wave.

Fig. 4. Photonic nanojets from vaterite and silica microparticles with diameters of $D_{particle}$=3 µm. (a), (b) Vaterite orientation-dependent nanojets. The nanojet lengths, imaged with the CCD camera, depend on the mutual orientation between the incident field polarization and the vaterite major axis, as verified numerically - (c) and (d), respectively. Dashed white lines in (a) and (b) denote the visual impression of the corresponding nanojets lengths. The incident wavevector $k_x$ in the inset in (c)-(d) shows the horizontal xy-projection of the total wavevector k. (e) The vaterite photonic nanojet field enhancement as the function of the vaterite major axis angle with respect to the incident wave polarization. Photonic nanojets are produced by the (f-h) vaterite and (i-k) silica particles. Nanoject bending towards a nearby particle (f), (h), (i), and (k). (g) and (j) - reference jets without a secondary particle present. The photographs (a)-(b), (f)-(k) were recorded in NIR. Numerical calculations are CW at 1.1 µm. Scale bar in (h) denotes 3 µm.

Fig. 5. The numerical analysis steps. (a) The principal scheme of the numerical analysis - FDTD model with two particles. The plane wave (CW, 1.1 µm wavelength) source has an incidence angle of 90°- α = 80°. (b) The calculated electromagnetic fields in Plane 1 (located beneath the particles in isotropic medium) are used for angular spectrum decomposition and consequent field calculations for a set of planes around the particles (i-th planes in (c)). Planes 2 and 3 in (a) are central cuts of the FDTD simulation box. To simulate the images on the CCD camera (i.e., Figs. 6(a-f)), a set of intensity distributions in i-th planes is averaged (see details in text).

Figure 6. Numerical analysis of inter-particle photonic nano-jet bending effect. (a), (b) Images of nanojets through the objective - "lens view", as elaborated in Fig.5. Vaterite particles with their central optical axes aligned with incident wavevector $k_x$ are used, as shown in the inset to (a). (c),(d) FDTD-calculated scattered near field intensity colormaps. No nanojet bending is found, in contrast to (a) and (b). (e),(f) The same as (a),(b), but with silica particles instead of vaterite ones. The particle's diameter in (a)-(f) is 3 µm. (g) Characteristic nanojet bending (Max $s_y$) as the function of the particle's refractive index. The mutual arrangement of representative particles is in the inset.

# Nanojet Visualization and Dark-field Imaging of Optically Trapped Vaterite Capsules with Endoscopic Illumination

Andrey Ushkov[1,2*], Andrey Machnev[1,2], Denis Kolchanov[1,2], Toms Salgals[3,4], Janis Alnis[5], Vjaceslavs Bobrovs[3], Pavel Ginzburg[1,2]

[1]Department of Electrical Engineering, Tel Aviv University, Ramat Aviv, Tel Aviv 69978, Israel
[2]Light-Matter Interaction Centre, Tel Aviv University, Tel Aviv, 69978, Israel
[3]Institute of Telecommunications, Riga Technical University, 12 Azenes Street, 1048 Riga, Latvia
[4]Nanophotonics research laboratory (NANO-Photon Lab.), Riga Technical University, Azenes street 12, LV-1048, Riga, Latvia
[5]Institute of Atomic Physics and Spectroscopy, University of Latvia, 3 Jelgavas Street, 1004 Riga, Latvia

## Supplementary material

*Darkfield imaging and jet visualization of several particles*

To illustrate the broader applicability of our method, we examined a variety of particles, encompassing different sizes and materials. Images of vaterite and $SiO_2$ particles appear in Fig.S1.

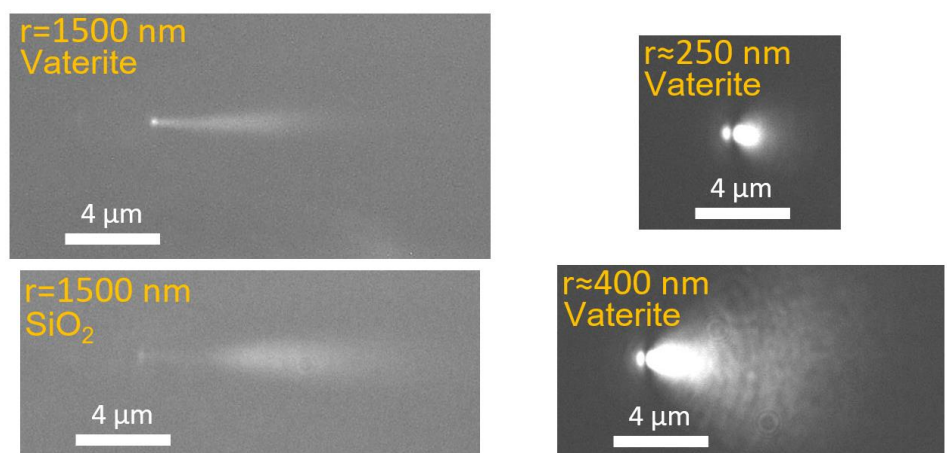

Fig.**S1**. Darkfield images of optically trapped particles of different sizes. The fiber endoscopic illumination comes from the left side. The nanojet pattern is clearly seen in all images, because of the high (~80°) incidence angle.

For the minimal radius of synthesized vaterite particles (250 nm), the optical setup demonstrates a strong darkfield signal. Fabricating vaterite particles of smaller sizes remains challenging.

*Conventional vs. Endoscopic Illumination – NA Analysis*

The quality of dark-field analysis is primarily determined by the amount of light that reaches the detector. This factor is, in turn, contingent on the numerical aperture of the collection optics. Here we will demonstrate that endoscopic illumination allows for collecting light from a broader NA. To estimate the improvement, we consider a 100 nm radius dielectric particle with refractive index n=1.6 (a mean of vaterite anisotropic indices of 1.55 and 1.65 for the sake of simplicity). Those parameters are relevant to nanomedical studies in the field, i.e., [1,2]. From an optical standpoint, such particles (and smaller) obey the Rayleigh regime, where only the electric non-resonant dipole contributes to scattering (see Fig.S2):

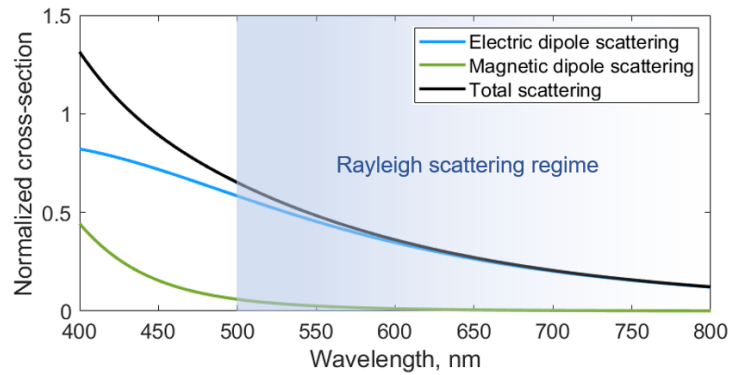

Fig.**S2**. The scattering cross-sections for a spherical dielectric particle (R=100 nm, n=1.6). The blue area denotes the spectral area, where the Rayleigh scattering regime occurs, and where electric dipole scattering is >10 times larger than magnetic dipole scattering.

Using Rayleigh approximation, we theoretically compare two optical darkfield setups: standard darkfield hollow-cone illumination, realized with a high-NA microscope objective, and our scheme with a fiber illumination, see Figure S3 for details:

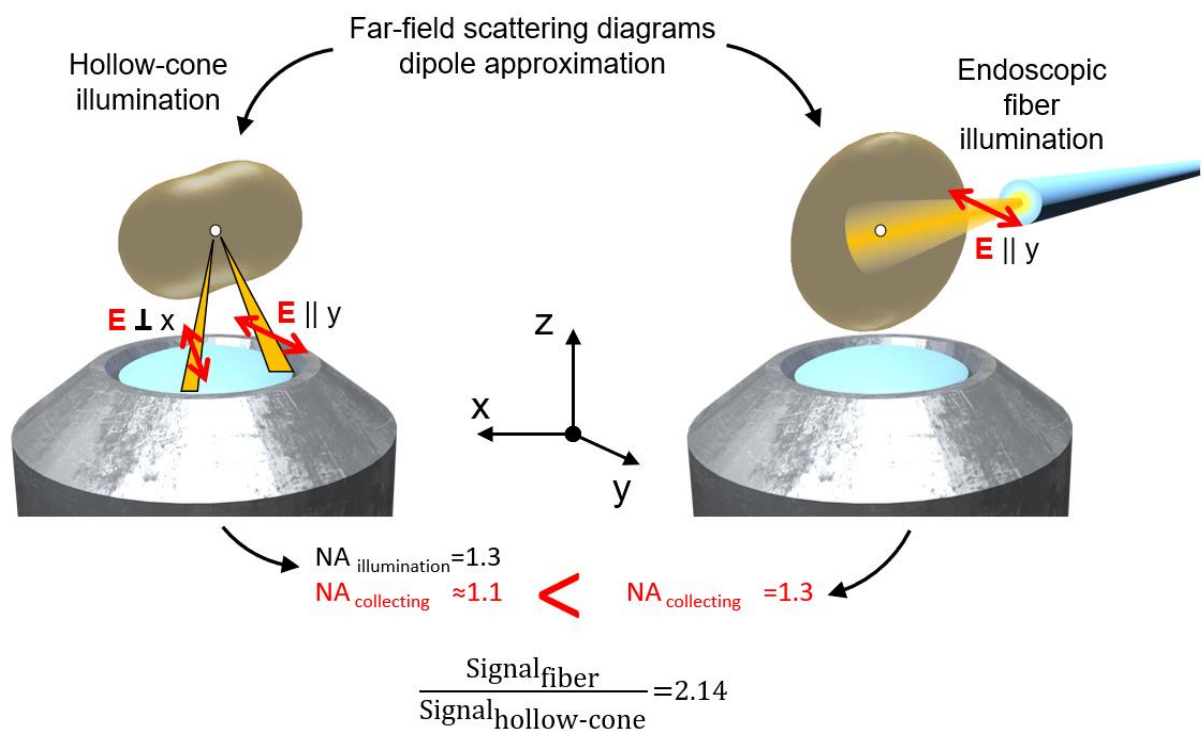

Fig.**S3**. The comparison of two darkfield illumination schemes: the standard hollow-cone illumination via the high-NA microscope objective (on the left) and our setup with a fiber illumination (on the right). Orange shapes around the particles (denoted as white dots above the objectives) represent the far-field scattering diagrams. The particles are immersed in ethanol (n=1.36).

We consider the monochromatic illumination at a wavelength > 500 nm. In this case, only electric dipoles oriented along the incident polarization are excited. In our setup, it leads to the single dipole excitation with a well-known toroidal shape scattering diagram in the far field (see Fig.S3), as the illumination has a well-defined linear polarization state. The hollow-cone illumination, however, leads to a more complex diagram because of the simultaneous excitation of multiple dipoles, corresponding to a span of azimuthal directions. *Note, that in the case of anisotropic particles, several modes can be exited, making this spectroscopic assessment less 'clean'.*

In a conventional dark-field setup, the objective plays the role of a condenser and refracts the light into the hollow-cone, which illuminates the particle. Consequently, this scheme cannot use its full NA=1.3 for the scattered light collection and an effective collecting $NA_{collecting} \approx 1.1$. In contrast, in our setup, the whole NA=1.3 is used solely for the scattered signal collection. For the sake of comparison, the incident light intensity in both cases is normalized to unity. The integration of the scattered signal into effective NA for both cases lead to the following intensity ratio: $I_{endoscopic-fiber}/I_{hollow-cone}=2.14$. In the calculation, the far-field scattering diagrams (different for both cases, as discussed above) are integrated within the collection angles. The estimate shows that a two-fold improvement is obtained owing to the endoscopic illumination.

*Estimating the nanojet bending capability*

The photonic nano-jet bending is the result of the near-field particle interaction, observed via the far-field mapping on the camera. The bending is visible both in experimental (Fig.4(f-k)) and numerical (Figs.6(a-b,e-f)) results when two particles are misaligned with respect to the incident field wavevector projection $\mathbf{k}_x$ on the horizontal plane (see, for example, the inset in Fig.6(a)). The particles and nanojet mutual arrangement is sketched in Figs.S4a (for the case of a straight nanojet) and Fig.S4c (for the case of a bent nanojet). When the particle offset $s_y$ changes, the nanojet tail (defined arbitrarily as a part of the nanojet in region $x>3D_{particle}$, where $D_{particle}=3$ µm is the particle diameter) tends to follow the misaligned particle (see Figs.S4a,c).

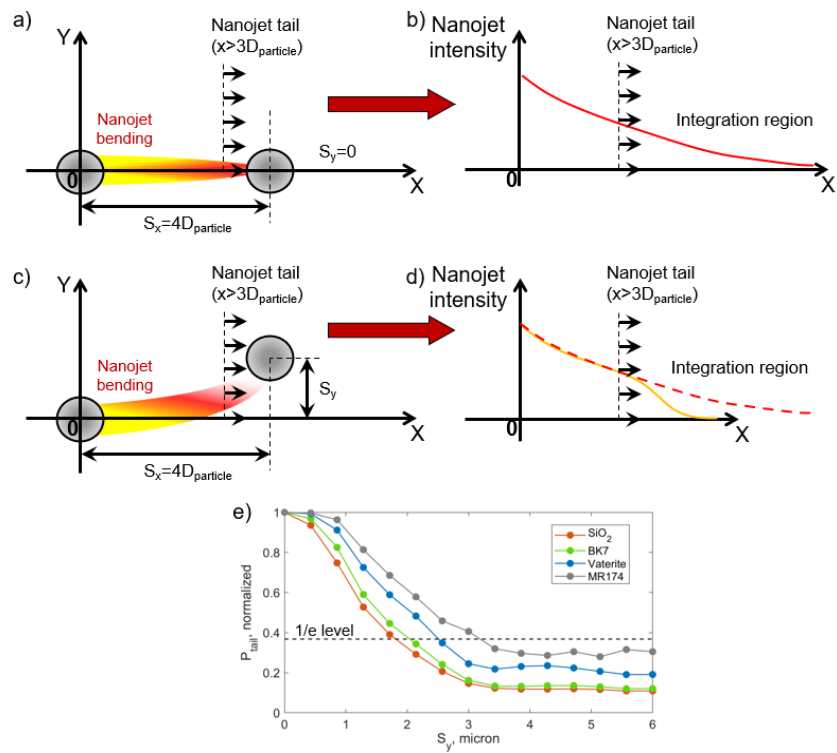

Fig.**S4**. (a),(c) The geometry of the nanojet formation. (b),(d) The nanojet intensity along the X-axis for two mutual arrangements of particles. Vertical lines demonstrate the transversal coordinate, where the intensity integration is made to obtain the value of $P_{tail}$. (e) The integrated nanojet tail power along the X-axis as a function of the secondary particle displacement. Color lines - different types of particles of the same 3 micron diameter. The maximum value of the particle offset $s_y$ is chosen at the 1/e level of the normalized power in the nanojet tail $P_{tail}$ (see details in the text).

To quantify the nanojet bending, the following approach has been used. For a set of $s_y$ values (from 0 µm to the arbitrarily chosen max value of $2D_{particle}$=6 µm) the nanojet intensity profiles along the X-axis were obtained, see Figs.S4b,d. Then, these intensity profiles were integrated within the tail region ($x>3D_{particle}$) only, to exclude the almost constant contribution of the brightest nanojet region near the particle boundary. The resulting curves $P_{tail}(s_y)$ are shown in Fig.S4e for a set of representative materials: $SiO_2$, BK7, vaterite, and MR174 (a high refractive index n=1.74 optical lens material). We characterize the nanojet bending by the value of the particle offset Max $s_y$, such that $P_{tail}(Max\ s_y)=P_{tail}(0)/e$, see Fig.6g of the main text.

**References**

[1] Sun, Ning, et al. "pH-dependent and cathepsin B activable CaCO3 nanoprobe for targeted in vivo tumor imaging." International Journal of Nanomedicine (2019): 4309-4317.
[2] Liu, Depeng, et al. "Oral delivery of insulin using CaCO3-based composite nanocarriers with hyaluronic acid coatings." Materials letters 188 (2017): 263-266.